\documentstyle[preprint,aps]{revtex}
\begin{document}
\def\a{\alpha}
\def\b{\beta}
\def\e{\epsilon}
\def\p{\partial}
\def\m{\mu}
\def\n{\nu}
\def\t{\tau}
\def\s{\sigma}
\def\g{\gamma}
\def\half{\frac{1}{2}}
\def\hatt{{\hat t}}
\def\hatx{{\hat x}}
\def\haty{{\hat y}}
\def\hatp{{\hat p}}
\def\hatX{{\hat X}}
\def\hatY{{\hat Y}}
\def\hatP{{\hat P}}
\def\hatth{{\hat \theta}}
\def\hatta{{\hat \tau}}
\def\hatrh{{\hat \rho}}
\def\hatva{{\hat \varphi}}
\def\p{\partial}
\def\nn{\nonumber}
\def\barx{{\bar x}}
\def\bary{{\bar y}}  
\def\cb{{\cal B}}
\def\2pap{2\pi\alpha^\prime}
\def\wideA{\widehat{A}}
\def\wideF{\widehat{F}}
\def\beq{\begin{eqnarray}}
\def\eeq{\end{eqnarray}}

\preprint{
KIAS-P01021
}

\title{Tachyon Condensation, Boundary State and\\
Noncommutative Solitons}

\author{{Taejin Lee}
\thanks{E-mail: taejin@cc.kangwon.ac.kr}}
\address{{\it 
    Department of Physics, Kangwon National University, 
    Chuncheon 200-701, Korea}}

\date{\today}
\maketitle
\begin{abstract}
We discuss the tachyon condensation in a single unstable D-brane in the
framework of boundary state formulation. The boundary state in the 
background of the tachyon condensation and the NS $B$-field is explicitly 
constructed. We show in both commutative theory and noncommutative
theory that the unstable D-branes behaves like an extended 
object and eventually reduces to the lower dimensional 
D-branes as the system approaches the infrared fixed point. We clarify the 
relationship between the commutative field theoretical description of the 
tachyon condensation and the noncommutative one.
\end{abstract}

\pacs{11.27, 11.25.-w, 11.25.Sq}

\narrowtext

\section{Introduction}

The fate of the unstable D-branes due to the tachyon condensation 
has been one of centers of interest in string theory since the seminal 
papers by Sen \cite{sen1}, where he conjectured that the unstable D-branes 
may behave like solitons corresponding to lower dimensional D-branes.
Since the tachyon condensation \cite{tach} is an off-shell phenomenon, we need to resort to 
the second quantized string theory in order to understand this noble phenomenon. 
There are two approaches available to this subject; the Witten's open string 
field theory \cite{witt} with the level truncation \cite{kost,senzw} and the 
boundary string field theory (BSFT) \cite{witt92,shata}. As the tachyon 
condensation occurs the open string field acquires an expectation value. 
In general it is quite difficult to solve the string field 
equation which involves an infinite number of components. 
One may solve the string field equation by truncating the string field to 
the first few string levels. This has been a useful practical tool to discuss the 
tachyon condensation. Many aspects of the tachyon condensation physics 
\cite{sfttr} were explored by this method. However, this method provides 
numerical results in most cases and sometimes yields qualitative ones only. 

The second approach, which is based 
on the background independent string field theory \cite{witt92}, has been 
developed recently \cite{shata} as an alternative tool to the string field 
theory with level truncation. 
It deals with the disk partition function of the open string theory, where the 
boundary of the disk is defined by the trace of the two ends of the open string. 
The configuration space for the BSFT is the space of two dimensional world-sheet 
theories on the disk with arbitrary boundary interactions. 
Endowing it with the antibracket structure in terms of the world-sheet path 
integral on the disk and the BRST operator, one can define $S$, the action of the 
target space theory. The relationship between the BSFT action, $S$ and the disk 
partition function $Z$ is clarified in ref.\cite{shata} as
\beq
S = \left( 1+ \b^i \frac{\p}{\p g^i} \right) Z
\eeq
where $g^i$ are the couplings of the boundary interactions and $\b^i$ are the 
corresponding world-sheet $\b$-functions. Choosing the tachyon profile, $T(X)$,
as for the boundary interaction, one can obtain the effective action for 
tachyon field \cite{eff}. Evaluation of the effective action may 
be simplified considerably if we introduce a large $NS$ B-field 
\cite{witten00,dasgupta,corbalba,oku,schnabl,harvey00}.
Recent works also show that $S$ may coincide with $Z$ if we 
introduce supersymmetry to describe the $D$-${\bar D}$ system 
\cite{kuta00,marino,niarchos}. 
The disk partition function $Z$ for the $D$-${\bar D}$ system has been 
explicitly evaluated in recent papers \cite{kuta00,kraus00}.

In this paper we discuss the tachyon condensation in the simplest 
setting, i.e., a single unstable D-brane in the bosonic string theory,
adopting the boundary state formulation, which developed by 
Callan, Lovelace, Nappi and Yost \cite{callan} sometime ago. 
The advantage of this approach is that one can explicitly construct the 
quantum states corresponding to the unstable D-brane systems. Hence, the 
couplings of the system to the various string states are readily obtained so 
it helps us to understand how the system evolves as the condensation occurs. 
This approach is closely related to the second one, the BSFT in that 
the normalization factor of the boundary state is simply the disk partition
function $Z$. Since the constructed boundary state is given as a 
quantum state of the closed string field, it may also help us to understand 
its relationship to the first approach based on the open string field theory,
if we appropriately utilize the open-closed string duality. 
We extend the boundary state formulation to the case of the 
noncommutative open string theory in order to discuss the 
noncommutative tachyon in the same framework. The relationship of the 
commutative theory and the noncommutative theory on the tachyon condensation 
can be understood in this framework along the line of the equivalence between
the commutative theory and the noncommutative theory of open string
\cite{seiberg,tlee}.

\section{Boundary Action and Boundary State}

We begin with the boundary state construction developed in ref.\cite{callan} 
and establish the relationship between the boundary action 
and the boundary state in more general cases. 
The boundary state formulation is based on a rather simple 
observation: It utilizes the open-closed string duality.
The disk diagram in the open string theory can be viewed 
equivalently as the disk diagram in the closed string theory.

\input epsf.tex
%
\epsfxsize=0.7\hsize
%
\epsfbox{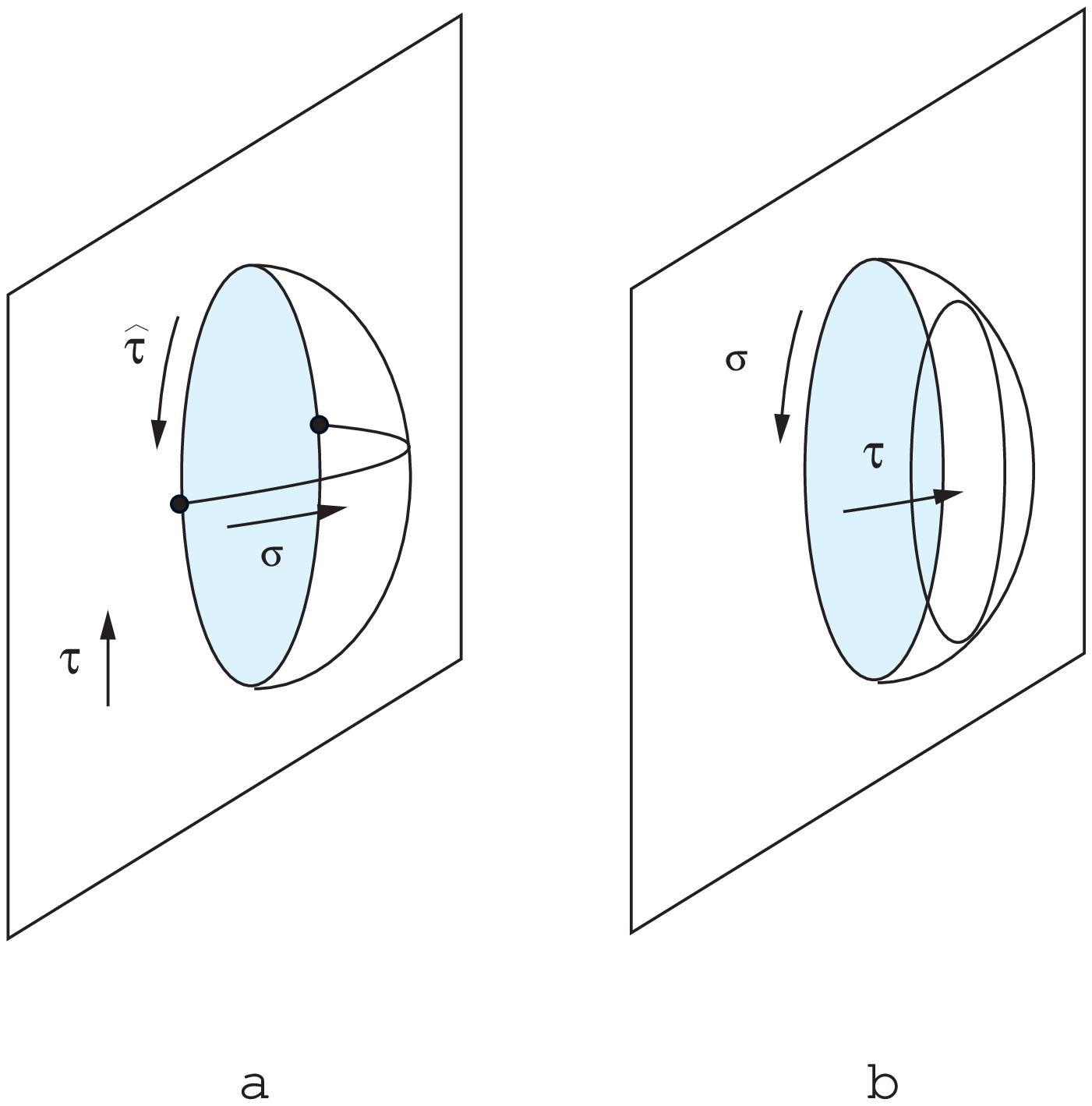}

\begin{center}
Figure 1: a. Disk diagram in open string theory,
b. Disk diagram in closed string theory
\end{center}

In the open 
string theory it depicts an open string appearing from the vacuum, 
then subsequently disappearing into the vacuum while in the 
closed string theory it depicts a closed string propagating from 
the boundary of the disk, then disappearing. This open-closed 
string duality is also useful when we deal with the cylindrical 
diagram. The cylindrical diagram make an appearance both in the 
open string theory and in the closed string theory. However, it 
can be interpreted differently in two theories. In the open string 
theory it describes a one-loop amplitude while in the closed 
string it describes a tree level amplitude. Since the closed 
string description often turns out to be simpler than its open string 
counterpart, the boundary state formulation has been employed as a practical 
method to evaluate the open string diagrams by making use of this 
open-closed string duality. It is proved to be extremely useful 
especially when we discuss various interactions between the 
$D$-brane and the open string.

We define the boundary state $\vert X \rangle$ by the 
following eigenvalue equation in the closed string theory
\beq  
\hatX^\m \vert X  \rangle = X^\m \vert X \rangle.  
\eeq 
Since $\hatX^\m$ and $X^\m$ are defined on $\p M$, the boundary of  
the disk, we may expand them as \cite{gsw}
\begin{mathletters}
\label{exp:all}
\beq 
\hatX^\m(\s) &=& \hatx^\m_0 + \sqrt{\frac{\a^\prime}{2}} \sum_{n=1} 
\frac{1}{\sqrt{n}} \left\{ (a^\m_n + \tilde{a}^{\m\dagger}_n) 
e^{2in\s} + (a^{\m\dagger}_n +\tilde{a}^\m_n) e^{-2in\s} \right\}
\label{exp:a}\\ 
X^\m(\s) &=& x^\m_0 + \sqrt{\frac{\a^\prime}{2}}
\sum_{n=1} \frac{1}{\sqrt{n}}\left(x^\m_n e^{2in\s} 
+ \barx^\m_n e^{-2in\s}\right) \label{exp:b} 
\eeq
\end{mathletters}
where $\s \in [0,\pi]$, $\m = 0, \dots, d-1$, and 
\beq 
a^\m_n = \frac{i}{\sqrt{n}} \a^\m_n, \quad a^{\m\dagger}_n =  
-\frac{i}{\sqrt{n}} \a^\m_{-n}, \quad 
\tilde{a}^\m_n = \frac{i}{\sqrt{n}} \tilde{\a}^\m_n, \quad  
\tilde{a}^{\m\dagger}_n = -\frac{i}{\sqrt{n}} \tilde{\a}^\m_{-n}.  
\nn
\eeq 
They satisfy the following commutation relationship
\beq
[a^\m_n, a^{\n \dagger}_m] = (g^{-1})^{\m\n} \delta_{nm},
\quad
[\tilde{a}^\m_n, \tilde{a}^{\n \dagger}_m] = (g^{-1})^{\m\n} 
\delta_{nm}.
\eeq
Hence the eigenvalue equation can be rewritten in terms of the left  
movers and the right movers as  
\beq 
\hatx^\m_0 \vert X \rangle &=& x^\m_0 \vert X \rangle \nn\\ 
\left(a^\m_n + \tilde{a}^{\m\dagger}_n \right) \vert X \rangle 
&=& x^\m_n \vert X \rangle \\ 
\left(a^{\m\dagger}_n + \tilde{a}^\m_n \right) \vert X \rangle 
&=& \barx^\m_n \vert X \rangle .\nn 
\eeq 
This eigenvalue equations determine the boundary state $\vert  
X\rangle = \vert x, \barx \rangle$ up to a normalization factor 
\beq 
\vert x, \barx \rangle = N(x, \barx) \prod_{n=1}  
\exp\left(-a^{\m\dagger}_n \tilde{a}^{\n\dagger}_n g_{\m\n} 
+ a^{\m\dagger}_n x^\n_n g_{\m\n}
+ \barx^\m_n \tilde{a}^{\n\dagger}_n g_{\m\n} \right) 
\vert 0 \rangle 
\eeq 
where $a_n \vert 0 \rangle = \tilde{a}_n \vert 0 \rangle = 0$. 
Requiring the completeness relation 
\beq 
\int D[x, \barx] |x, \barx \rangle \langle x, \barx | = I, \label{comp}
\eeq 
we may fix the normalization factor 
\beq 
\vert x, \barx \rangle = \prod_{n=1}\exp \left\{ -\half \barx_n  
x_n -a^\dagger_n\tilde{a}^\dagger_n +a^\dagger_n x_n+ 
\barx_n\tilde{a}^\dagger_n\right\} \vert 0\rangle 
\label{xx}
\eeq 
where the space-time indices are suppressed and the contraction 
with the metric $g_{\m\n}$ is implied.
 
We will make use of the set of the boundary states,  
$\{ \vert x, \barx \rangle \}$ as a basis to construct various  
boundary states, which correspond to the disk diagrams with nontrivial  
backgrounds: For a given boundary action $S_{\p M}$, the boundary  
state is defined as 
\beq 
\vert B \rangle = \int D[x, \barx] e^{iS_{\p M}[x, \barx]} 
\vert x, \barx  \rangle . \label{bound} 
\eeq 
Here $S_{\p M}[x, \barx]$ is the boundary action evaluated with  
the boundary condition 
\beq 
X^\m \vert_{\p M} &=& x^\m_0 + \sqrt{\frac{\a^\prime}{2}}
\sum_{n=1} \frac{1}{\sqrt{n}}\left(x^\m_n e^{2in\s} 
+ \barx^\m_n e^{-2in\s}\right). \label{bc1} 
\eeq 
The requirement of the completeness relation Eq.(\ref{comp}) and the
definition  of the boundary state, Eq.(\ref{bound}) are consistent with the
closed string field theory. The boundary state in fact defines a 
quantum state of the closed string field. 
For a free string, $S_{\p M} = 0$, 
\beq 
|B_{Free} \rangle = \int D[x, \barx]  \vert x, \barx  
\rangle = \sqrt{\det g} \prod_{n=1} \exp\left(a^{\m\dagger}_n 
\tilde{a}^{\n\dagger}_n g_{\m\n} \right) \vert 0 \rangle, 
\eeq 
which satisfies the boundary condition 
\beq 
a^{\m\dagger}_n |B_{Free} \rangle = \tilde{a}^\m_n |B_{Free} \rangle,  
\quad 
\tilde{a}^{\m\dagger}_n |B_{Free} \rangle = a^\m_n |B_{Free} \rangle. 
\eeq 
This boundary condition, of course, nothing but the Neumann  
boundary condition, $\p_\t X^\m \vert_{\p M} = 0$. The boundary state,
satisfying the Dirichlet condition $\hatX^\m |B \rangle = 0$,
is 
simply obtained by taking $x^\m = x^\m_n = {\bar x}^\m_n = 0$ in
Eq.(\ref{xx})
\beq
|B_{Dirichlet} \rangle = \sqrt{\det g} \prod_{n=1}\exp 
\left(-a^{\m\dagger}_n \tilde{a}^{\n\dagger}_n g_{\m\n} \right) 
\vert 0\rangle. 
\eeq
It follows that the boundary state corresponding to a flat D-p-brane 
is given as 
\beq
|Dp \rangle = \frac{T_p}{g_s} \sqrt{\det g} \prod_{n=1} 
\exp\left(a^{i\dagger}_n \tilde{a}^{j\dagger}_n g_{ij}
-a^{a\dagger}_n \tilde{a}^{b\dagger}_n g_{ab} \right)|0 \rangle
\eeq
where $i = 0, \dots, p$, $a = p+1, \dots, d-1$. Here $T_p$ and 
$g_s$ are the tension of the $Dp$-brane and the string coupling
constant respectively.

One of the well-known examples of the boundary state is the open 
string in a constant $U(1)$ background. The $U(1)$ background yields the  
following boundary action for an open string
\beq 
S_{F} =  \int_{\p M} d\hat{\t} A_\m \frac{\p X^\m}{\p \hat{\t}}
= \frac{1}{2} \int_{\p M} d\hat{\t} F_{\m\n} X^\m 
\frac{\p X^\n}{\p \hat{\t}} \label{bcact}
\eeq 
where ${\hat \tau}$ is a proper-time parameter along $\p M$
\beq
{\hat \tau} =
\left\{\begin{array}{r@{\quad:\quad}l}
\tau-1 &  {\hat \tau} \in [-1,0] \\
-\tau+1 & {\hat \tau} \in [0,1]  .
\end{array} \right. \label{hatta}
\eeq
In the closed string world-sheet coordinates the boundary 
interaction reads as 
\beq
S_F = \frac{1}{2} \int_{\p M} d\s F_{\m\n} X^\m 
\frac{\p X^\n}{\p \s}.
\eeq
The boundary interaction Eq.(\ref{bcact}) yields the boundary condition for the
open string as 
\beq
\frac{1}{\2pap} g_{\m\n} \p_\s X^\n - F_{\m\n} \p_\t X^\n = 0, 
\quad {\rm on} \,\,\, \p M.
\eeq
In the closed string world-sheet coordinates we get 
the boundary condition as 
\beq
\frac{1}{\2pap} g_{\m\n} \p_\t X^\n - F_{\m\n} \p_\s X^\n = 0, 
\quad {\rm on} \,\,\, \p M.
\eeq
We may transcribe it into the boundary condition to be imposed on
the boundary state
\begin{mathletters}
\label{bc:all}
\beq
a_n |B_F \rangle &=& (g + \2pap F)^{-1} (g - \2pap F) 
\tilde{a}^\dagger_n |B_F \rangle, \label{bc:a} \\
\tilde{a}_n |B_F \rangle &=& (g - \2pap F)^{-1} (g + \2pap F) 
a^\dagger_n |B_F \rangle . \label{bc:b}
\eeq
\end{mathletters}
making use of the closed string mode expansion of $\hatX^\m (\t,\s)$
\beq
\hatX^\m(\t,\s) &=& \hatx^\m +2\a^\prime p^\m \t+ 
i\sqrt{\frac{\a^\prime}{2}} \sum_{n \not= 0} \frac{1}{n} \left(
\a^\m_n e^{-2in(\t-\s)} + \tilde{\a}^\m_n e^{-2in(\t+\s)} \right) 
\nn \\ &=& \hatx^\m +2\a^\prime p^\m \t+ 
\sqrt{\frac{\a^\prime}{2}} \sum_{n=1} \frac{1}{\sqrt{n}} (
a^\m_n e^{-2in(\t-\s)} + a^{\m\dagger}_n e^{2in(\t-\s)} \nn\\
& & \quad + \tilde{a}^\m_n e^{-2in(\t+\s)} + \tilde{a}^{\m\dagger}_n 
e^{2in(\t+\s)}). \label{expnd}
\eeq

With the given boundary condition Eq.(\ref{bc1}), the boundary action 
is evaluated as 
\beq 
S_F = (\2pap) \frac{i}{2} \sum_{n=1} \barx^\m_n F_{\m\n} x^\n_n .
\eeq 
Then a simple algebra Eq.(\ref{bound}) leads us to the boundary state 
$\vert B_F \rangle$ 
\beq 
|B_F \rangle = \frac{T_p}{g_s} \prod_{n=1} \det \left(g +\2pap F \right)^{-1} 
\exp \left\{a^\dagger_n g \left(g+ \2pap F\right)^{-1}
\left(g- \2pap F\right) \tilde{a}^\dagger_n \right\}\vert 0 \rangle 
\eeq 
which satisfies the desired boundary condition Eq.(\ref{bc:all}).
It should be noted that the normalization factor of the  
boundary state is the well-known Dirac-Born-Infeld Lagrangian 
\beq 
Z =\frac{T_p}{g_s} \prod_{n=1} \det \left(g + \2pap F \right)^{-1} = 
\frac{T_p}{g_s} \sqrt{\det(g + \2pap F)}, 
\eeq 
where we make use $\zeta(0) = \sum_{n=1} 1 = -/2$.
It can be also obtained by evaluating the Polyakov string path integral  
on a disk \cite{frad}.  
 
The relationship between the boundary action and the boundary state
observed in the case of the $U(1)$ background can be established in 
more general cases. In order to see this explicitly let us introduce 
a boundary action of more general form as  
\beq 
S &=& S_M + S_{\p M}, \label{action} \\ 
&=& - \frac{1}{4\pi \a^\prime} \int_M d\t d\s \sqrt{-h} h^{\a\b} g_{\m\n} 
\p_\a X^\m \p_\b X^\n + \frac{i}{2} \int_{\p M} d\s X^\m M_{\m\n} X^\n. \nn 
\eeq 
Here $M_{\m\n} = M_{\m\n} \left(\frac{1}{i} \frac{\p}{\p \s}\right)$
is a differential/integral operator in $\s$.
From this action we get a bulk equation on $M$ as usual
\beq 
(\p^2_\t - \p^2_\s) X^\m = 0, \label{eqm} 
\eeq 
and a boundary condition to be imposed on $\p M$ 
\beq 
\frac{1}{\2pap} g_{\m\n} \p_\t X^\n - iM_{\m\n}
\left(\frac{1}{i} \frac{\p}{\p \s}\right) X^\n =0. 
\eeq 
Making use of the mode expansion of $\hatX^\m$, Eq.(\ref{expnd}),
we may transcribe the boundary condition into operator equations acting on the  
boundary state as 
\beq 
\hatp^\m |B \rangle &=& i \pi (g^{-1}M)^\m{}_\n \hatx^\n | B \rangle, \nn\\ 
a^\n_n |B \rangle &=& \left(g + \frac{\pi\a^\prime}{n} M(2n) \right)^{-1} 
\left(g - \frac{\pi\a^\prime}{n} M(2n) \right) \tilde{a}^{\n\dagger}_n |B  
\rangle, \label{bcmode}\\ 
\tilde{a}^\n_n |B \rangle &=& \left(g + \frac{\pi\a^\prime}{n} M(-2n) 
\right)^{-1} \left(g - \frac{\pi\a^\prime}{n} M(-2n) \right) 
a^{\n\dagger}_n |B \rangle. \nn 
\eeq 
From the action we see that 
\beq 
M(2n)^T = M(-2n), 
\eeq 
which ensures consistency of our construction. 
 
Given the boundary conditions Eq.(\ref{bcmode}) one can determine the  
boundary state, but only up to a normalization factor. Since the  
normalization factor is also often important, we should find a way  
to fix it. To this end one may calculate couplings of the system to  
the closed string degrees of freedom. Interpreting them as linear  
variations of the effective action, which is supposed to be obtained as 
the disk partition function, one may fix the normalization factor. 
This procedure has been applied to the open string in the constant 
$U(1)$ background in ref. \cite{vecch}. However, it would be involved 
in more general cases. 
Here, in order to fix the normalization factor we simply take 
Eq.(\ref{bound}) as the definition of the boundary state. 
With the given boundary condition Eq.(\ref{bc1}) we find that the 
boundary action $S_{\p M}[x,\barx]$ is obtained as 
\beq 
S_{\p M} = i \frac{\pi}{2}  x^\m M_{\m\n}(0) x^\n +i \frac{\pi\a^\prime}{2} 
\sum_{n=1} \frac{1}{n} \barx_n^\m M(2n)_{\m\n} x^\n_n. 
\eeq 
Then a Gaussian integral of Eq.(\ref{bound}) brings us to 
an explicit expression of the boundary state  
\beq 
|B \rangle &=& Z_{Disk} \exp \left\{ a^\dagger_n g  
\left(g+ \frac{\pi\a^\prime}{n} M(2n) \right)^{-1} 
\left(g - \frac{\pi\a^\prime}{n} M(2n) \right)
\tilde{a}^\dagger_n \right\} | 0\rangle, \label{boundstate}\\ 
Z_{Disk} &=& \frac{T_p}{g_s} \frac{1}{\sqrt{\det M(0)}} \prod_{n=1} 
\det\left(g+ \frac{\pi\a^\prime}{n} M(2n) \right)^{-1} \nn  
\eeq 
Here, $Z_{Disk}$ is the disk partition function with the  
boundary action $S_{\p M}$. It is easy to see that the boundary state  
$|B \rangle$ readily satisfies the desired boundary condition, 
Eq(\ref{bcmode}). Eq.(\ref{boundstate}) exhibits clearly the relationships 
between the normalization factor, the disk partition function and 
the boundary condition in the framework of boundary state formulation. 
 
\section{Tachyon Condensation and Boundary State}

Being equipped with the boundary state formulation given in the previous
section, we construct the boundary state in the background of the tachyon
condensation and discuss some important physical properties of the unstable
D-brane systems. The tachyon condensation introduces the following boundary  
interaction in the closed string world-sheet coordinates
\beq 
S_{T} = i \int_{\p M} d\s T(X). 
\eeq 
In order to construct the boundary state explicitly for the string in the
tachyon background we consider a simple tachyon profile, 
$T(X) = u_{\m\n} X^\m X^\n$.   
In terms of the normal modes $S_T$ is written as 
\beq 
S_T =i\pi x^\m u_{\m\n} x^\n + i \pi \a^\prime \sum_{n=1} \frac{1}{n}  
\barx^\m_n u_{\m\n} x^\n_n 
\eeq 
With this boundary action we construct the boundary state, 
following Eq.(\ref{bound}), 
\beq 
|B_T \rangle &=& \int D[x, \barx] e^{i S_T} \vert x, \barx  
\rangle \nn \\ &=& \int D[x, \barx] e^{-\pi x u x }  
\exp\left\{-\half \barx_n \left(g+ \frac{\2pap}{n}u\right)x_n 
- a^\dagger_n g \tilde{a}^\dagger_n + a^\dagger_n g x_n + 
\barx_n g \tilde{a}^\dagger_n  \right\} \vert 0\rangle \nn\\ 
&=& Z_{Disk} \exp\left\{a^\dagger_n g \left(g+\frac{\2pap}{n} u 
\right)^{-1} \left(g- \frac{\2pap}{n} u \right)\tilde{a}^\dagger_n 
\right\} |0\rangle,  \\
Z_{Disk} &=& \frac{T_p}{g_s} \frac{1}{\sqrt{\det(u)}}\prod_{n=1} 
\det \left(g+ \frac{\2pap}{n}u \right)^{-1} \nn
\eeq 
where we suppress the space-time indices.
It is easy to confirm that this boundary state satisfies the  
appropriate boundary condition 
\beq 
\left(\frac{1}{\2pap} g_{\m\n} \p_\t \hatX^\n -2i u_{\m\n} 
\hatX^\n \right)|B_T \rangle= 0. \eeq 
We note that the normalization factor $Z_{Disk}$ coincides with the disk 
amplitude for the open string in the background of tachyon condensation
as expected.

Now let us introduce the $U(1)$ background with a constant $F$ in addition to 
$S_T$, i.e., the boundary action $S_{\p M}$ is given as 
\beq
S_{\p M} = S_T + S_F,
\eeq
which corresponds to the case where 
\beq
M(0) = 2u, \quad M(2n) = 2n F + u, \quad n \ge 1.
\eeq
Accordingly the boundary state is constructed to be 
\beq
\vert B_{F+T} \rangle &=& \int D[x, \barx] e^{iS_F+iS_T} \vert x, \barx 
\rangle \nn\\
&=& Z_{Disk} \exp\left\{a^\dagger_n g \left(g+ \2pap F+ 
\2pap\frac{u}{n} \right)^{-1} \left(g- \2pap F -\2pap\frac{u}{n} \right)
\tilde{a}^\dagger_n \right\} |0\rangle, \\
Z_{Disk} &=& \frac{T_p}{g_s} \frac{1}{\sqrt{\det(u)}}\prod_{n=1} 
\det \left(g+ \2pap F+ \2pap\frac{u}{n} \right)^{-1}. \nn
\eeq
The boundary state $\vert B_{F+T} \rangle$ satisfies the following
boundary condition
\beq
\left(g_{\m\n} \p_\t \hatX^\n - \2pap F_{\m\n} \p_\s \hatX^\n -
4\pi\a^\prime i u_{\m\n}\hatX^\n \right) \vert B_{F+T} \rangle = 0.
\eeq
When $F$ is skew-diagonal with $F_{2\m-1, 2\m} = f_\m$, $u$ is diagonal,
$u_{\m\n} = u_\m \delta_{\m\n}$, $g_{\m\n} = \delta_{\m\n}$, and $\2pap =1$,  
the normalization factor reduces to 
\cite{andreev}
\beq
Z_{Disk} = \frac{T_p}{g_s} \frac{1}{\sqrt{\det u}} 
\prod_{n=1}\prod_{\m =1}^{d/2}\left\{\left(1+ \frac{u_{2\m-1}}{n}\right) 
\left(1+ \frac{u_{2\m}}{n}\right) +f^2_\m \right\}^{-1}.
\eeq 
If we are concerned the unstable $Dp$-brane in $d$ dimensions, 
\beq
F_{ab} =F_{ai} = F_{ia} =0, \quad u_{ab} = u_{ai} = u_{ia} =0,\nn
\eeq
where $i$, $j = 0, \dots, p$ and $a$, $b = p+1, \dots, d-1$, thus, 
\beq
\vert B_{F+T} \rangle &=& Z_{Disk} \prod_{n=1} \exp\left\{a^\dagger_n g 
\left(g+ \2pap F+ \2pap\frac{u}{n} \right)^{-1} \left(g- \2pap F 
-\2pap\frac{u}{n} \right) \tilde{a}^\dagger_n \right\} \nn\\
& & \exp \left(-a^{a\dagger}_n \tilde{a}^{b\dagger}_n g_{ab} \right) 
|0\rangle, \label{dpbrane}\\
Z_{Disk} &=& \frac{T_p}{g_s}\frac{1}{\sqrt{\det(u)}}\prod_{n=1} \det \left(g+ \2pap F+ 
\2pap\frac{u}{n} \right)^{-1}. \nn
\eeq
In Eq.(\ref{dpbrane}) $g$, $u$ and $F$ are $(p+1) \times (p+1)$ 
matrices.

Here one can make a simple observation on the effect of the tachyon 
condensation on the boundary state wavefunction. In order to see 
it we may leave the integration over $x$,
\beq
|B \rangle = \int dx |B; x \rangle.
\eeq
Then we find 
\beq
|B; x\rangle \sim e^{-\pi x u x}.
\eeq
It implies that the spatial dimension of the system is order of $1/\sqrt{\det u}$
if we use a closed string as a probe. 
The unstable D-brane may behave like a soliton in the lower 
dimensions and it becomes sharply localized as the system approaches the 
infrared fixed point, $u \rightarrow \infty$. 
One may be concerned about its behaviour at the infrared fixed
point since the boundary state may become singular as the system is 
sharply localized. In order to examine the behaviour of a $Dp$-brane near the 
infrared fixed point, let us suppose that tachyon condensation takes place only
in one direction, i.e., $u_{ij} = u_{ip} = u_{pi} = 0$, $i, j = 0, 
\dots, p-1$. Then as $u_{pp}=u \rightarrow \infty$, 
\beq
Z_{Disk} &=& \lim_{u \rightarrow \infty} \frac{T_p}{g_s} \int d^{p+1} x 
e^{-\pi u x^2_p} \prod_{n=1} \det\left[g+\2pap F+ \2pap u/n\right]^{-1}_{(p+1) \times 
(p+1)} \nn\\
&=& \lim_{u \rightarrow \infty} \frac{T_p}{g_s} \int d^{p} x \frac{1}{\sqrt{u}}
\prod_{n=1} \det\left[g+\2pap F \right]^{-1}_{p \times p}
\frac{1}{\2pap} \left(\frac{u}{n} \right)^{-1} \\
&=& 2\pi\sqrt{\a^\prime} \frac{T_p}{g_s} \int d^p x 
\sqrt{\det\left[g+\2pap F \right]}_{p \times p} \nn
\eeq
where $[A]_{m \times m}$ denotes a $m \times m$ matrix
and the zeta function regularization is used \cite{kraus00}
\beq
\prod_{n=1} \frac{1}{n +\e} = \exp \left\{
\frac{d}{ds} \left(\zeta(s, \e) - \e^{-s}\right) \right\}_{s=0} 
= \frac{\e \Gamma(\e)}{\sqrt{2\pi}}.
\eeq
This is precisely the disk partition function for a $(p-1)$ 
dimensional D-brane in the $U(1)$ background. 
Note also it gives us the correct 
relationship between the tension of a $Dp$-brane and that of a 
$D(p-1)$-brane
\beq
T_{p-1} = 2\pi \sqrt{\a^\prime} T_p
\eeq
Accordingly as $u \rightarrow \infty$,
\beq
\vert B_{F+T} \rangle &=& \frac{T_{p-1}}{g_s} \int d^{p} x 
\sqrt{\det[g+\2pap F]}_{p \times p} \nn\\
& & \prod_{n=1} \exp\left\{a^{\dagger i}_n \left[g(g+ \2pap F)^{-1}
(g -\2pap F) \right]_{ij}\tilde{a}^{\dagger j}_n -
a^{\dagger a}_n \tilde{a}^{\dagger b}_n g_{ab}
\right\} |0\rangle \label{f+t}
\eeq
where $i, j = 1, \dots, p-1$ and $a, b = p, \dots, d-1$. 
The first term in the exponent of Eq.(\ref{f+t}) 
describes the boundary state corresponding to an open string in 
$(p-1+1)$ dimensions with a constant $U(1)$ background and the second term 
implies that the boundary conditions along the directions 
$(x^p, \dots, x^{d-1})$ are Dirichlet. 
Thus, the unstable D-brane turns into the low dimensional D-brane at the 
infrared fixed point. As the system reaches the infrared fixed point, 
it becomes sharply localized. And as we may expect,
its wavefunction gets a divergent contribution from the zero mode, but
it is cancelled by the those from the higher modes. The cancellation
occurs only when we include contributions of all higher modes. 
This phenomenon is also observed in the $D$-${\bar D}$ system 
\cite{kuta00,kraus00,alwis}.

Since the boundary state in the background of the tachyon condensation is
explicitly constructed, the couplings to the closed string states are readily
obtained. The massless closed string states are given in the 
bosonic string theory as 
\beq
e_{\m\n} a^{\dagger \m}_1 \tilde{a}^{\dagger \n}_1 | k \rangle
\eeq
where $\m, \n = 0, \dots, d-1$.
Here $e_{\m\n}$ is chosen as 
\beq
e_{\m\n} = h_{\m\n}, \quad h_{\m\n} = h_{\n\m}, \quad k^\m h_{\m\n} = 
\eta^{\m\n} h_{\m\n} = 0
\eeq
for the graviton,
\beq
e_{\m\n} = \frac{\phi}{2\sqrt{2}} \left(\eta_{\m\n} -k_\m l_\n -k_\n l_\m 
\right), \quad l^2=0, \quad k \cdot l =1
\eeq
for the dilaton, and
\beq
e_{\m\n} = \frac{1}{\sqrt{2}} \Lambda_{\m\n}, \quad \Lambda_{\m\n} = 
-\Lambda_{\n\m}, \quad k^\m \Lambda_{\m\n} = 0
\eeq
for the Kalb-Ramond field.
(It may be more appropriate to discuss the couplings to the closed 
string states in the super-string theory. But the general features 
of the couplings in the tachyon background discussed here remain 
unchanged.)
The coupling of the boundary state to the massless closed string 
state is given as 
\beq
\langle k |e_{\m\n} a^\m_1 \tilde{a}^\n_1 |B \rangle &=&
\langle k |e_{ij} a^i_1 \tilde{a}^j_1 |B \rangle
+ \langle k |e_{ab} a^a_1 \tilde{a}^b_1 |B \rangle \nn\\
&=& e_{ij} \left[\left(g+ \2pap F+ \2pap \frac{u}{n} 
\right)^{-1} \left(g- \2pap F -\2pap \frac{u}{n} \right)g^{-1} 
\right]^{ji} Z \\
&& - e_{ab} (g^{-1})^{ab} Z \nn
\eeq
where $i=0, 1, \dots, p$ and $a= p+1, \dots, d-1$.
Here the boundary state $|B \rangle$ is given by Eq.(\ref{dpbrane}).
(An improved form of the coupling of the 
boundary state to the massless closed string state has been 
discussed recently in ref.\cite{semenoff}.)
Let us suppose that $u_{pp} = u \rightarrow \infty$, then
\beq
| B\rangle \rightarrow | B^\prime \rangle
&=& Z_{Disk} \prod_{n=1} \exp \Biggl\{ a^{\dagger i^\prime}_n 
\Bigl[g\left(g+ \2pap F+ \2pap \frac{u}{n} \right)^{-1} \nn\\
& & \left(g- \2pap F -\2pap \frac{u}{n} \right) 
\Bigr]_{i^\prime j^\prime} \tilde{a}^{\dagger j^\prime}_n
\quad  - a^{\dagger a^\prime}_n \tilde{a}^{\dagger a^\prime}_n 
\Biggr\}| 0\rangle \\
Z_{Disk} &=& \frac{T_p}{g_s} \frac{1}{\sqrt{\det [u]}_{p \times p}} 
\prod_{n=1} \det \left(
g+ \2pap F + \2pap \frac{u}{n} \right)^{-1}_{p \times p} \nn
\eeq
and 
\beq
\langle k | e_{\m\n} a^\m_1 \tilde{a}^\n_1 |B \rangle \rightarrow
\langle k | e_{\m\n} a^\m_1 \tilde{a}^\n_1 |B^\prime \rangle
&=& e_{i^\prime j^\prime} \Bigl[\Bigl(g+ \2pap F+ \2pap \frac{u}{n} 
\Bigr)^{-1} \nn \\
& & \Bigl(g- \2pap F -\2pap \frac{u}{n} \Bigr)g^{-1} 
\Bigr]^{j^\prime i^\prime}Z- e_{a^\prime b^\prime} 
(g^{-1})^{a^\prime b^\prime}Z 
\eeq
where $i^\prime, j^\prime = 0, 1, \dots, p-1$, and $a^\prime, 
b^\prime = p, \dots, d-1$.
This is exactly the coupling of the boundary state in lower 
dimensions to a massless closed string state. 
At a glance one can realize that the couplings of the massive 
closed string states also turn into the couplings to the lower
dimensional D-brane as the system reaches the infrared fixed point.
Thus, if we use a closed string as a probe, the unstable D-brane
looks identical with the lower dimensional D-brane at the infrared
fixed point.

\section{Tachyon Condensation and Noncommutative Solitons}

In recent papers \cite{witten00,dasgupta,corbalba,oku,schnabl,harvey00}
it has been pointed out that the tachyon 
condensation may be greatly simplified if one introduces a large NS 
$B$-field on the world-sheet. Some properties of the unstable systems, 
such as the tachyon potential and the D-brane tension, can be calculated 
exactly. Here in this section we will discuss the noncommutative 
tachyon in the same framework of the boundary state formulation. 
We may recall that in the canonical quantization \cite{tlee} 
one can establish the equivalence between the noncommutative 
open string theory with the commutative one in the presence of the NS 
$B$-field background. Along this line we can establish the 
relationship between the noncommutative tachyon theory and 
the commutative one. 

The bosonic part of the classical action for an open string ending on 
a $Dp$-brane with a NS $B$-field is given by 
\beq
S_M + S_B = -\frac{1}{4\pi\a^\prime} \int_M d^2 \xi \left[
g_{\m\n} \sqrt{-h} h^{\a\b} \frac{\p X^\m}{\p \xi^\a}
\frac{\p X^\n}{\p \xi^\b} - \2pap B_{ij} e^{\a\b}
\frac{\p X^i}{\p \xi^\a} \frac{\p X^j}{\p \xi^\b} \right]
\eeq
where $\m, \n= 0, 1, \dots, d-1$ and $i, j=0, 1, \dots, p$.
Let us consider the commutative description first. With a constant 
$B$ the second term yields the boundary action
\beq
S_B = \frac{1}{2} \int_{\p M} d\t B_{ij} X^i \p_\t X^j.
\eeq
As we transform the open string world-sheet coordinates to the 
closed string world-sheet coordinates, the boundary action turns 
into
\beq
S_B = \frac{1}{2} \int_{\p M} d\s B_{ij} X^i \p_\s X^j,
\eeq
which is of the same form as the constant $U(1)$ background discussed in 
the previous section. Hence, we get the corresponding boundary state in 
the background of tachyon condensation by replacing $F$ with $B$ in  
Eq.(\ref{dpbrane})
\beq \label{bcbt}
\vert B_{B+T} \rangle &=& Z_{Disk} \prod_{n=1} 
\exp\left\{a^{i\dagger}_n (g{\cal M}_{B+T})_{ij} \tilde{a}^{j \dagger}_n
-a^{a\dagger}_n \tilde{a}^{b\dagger}_n g_{ab} \right\} 
|0\rangle, \label{cbc}\\
({\cal M}_{B+T}) &=& \left(g+ \2pap B+ \2pap\frac{u}{n} \right)^{-1} \left(g- \2pap B 
-\2pap\frac{u}{n} \right), \nn\\
Z_{Disk} &=& \frac{T_p}{g_s} \frac{1}{\sqrt{\det(u)}}\prod_{n=1} 
\det \left(g+ \2pap B+ \2pap\frac{u}{n} \right)^{-1} \nn
\eeq
where $i, j = 0, \dots, p$ and $a, b = p+1, \dots, d-1$.
Since we are interested to compare the commutative description 
with the noncommutative one, we consider the case where $Dp$-brane 
$\rightarrow$ $D(p-2)$-brane. In this case we take 
\beq
u_{pp} = u_{p-1,p-1} = u \rightarrow \infty, \quad
u_{ij} = B_{ij} = 0,
\eeq
for $i, j = 0, \dots, p-2$ and the boundary state describes the 
$D(p-2)$-brane
\begin{mathletters}
\beq
\vert B_{B+T} \rangle &=& Z_{Disk} \prod_{n=1} \exp\left\{a^{i \dagger}_n 
g_{i j} \tilde{a}^{j \dagger}_n -a^{a\dagger}_n \tilde{a}^{b \dagger}_n 
g_{ab} \right\} |0\rangle \\
Z_{Disk} &=& \frac{T_{p-2}}{g_s} \sqrt{\det g}
\eeq
\end{mathletters}
where $i, j = 0, \dots, p-2$ and $a, b = p-1, \dots, d-1$. 
Note that the role of the NS $B$-field is rather trivial in the 
commutative description of the tachyon condensation. 

Let us turn to the noncommutative description. Since the NS 
B-field term is quadratic in string variables, $X$, one may define the 
world-sheet Green function with respect to $S_M + S_B$ instead of 
$S_M$. Then it leads us to the noncommutative open string theory, 
of which world-sheet Hamiltonian and the string variables in 
the longitudinal directions are given 
as \cite{tlee}
\begin{mathletters}
\label{str:all}
\beq
H &=& (2\pi\a^\prime) \half p_i (G^{-1})^{ij} p_j
+(2\pi\a^\prime) \sum_{n=1}\left\{ \half K_{ni} (G^{-1})^{ij}
K_{nj} + \frac{1}{(2\pi\a^\prime)^2} \frac{n^2}{2} Y_{ni} G_{ij} Y_{nj} 
\right\}, \label{str:a}\\
X^i(\s) &=& x^i+i \theta^{ij}p_j \left(\s - \frac{\pi}{2}\right)+
\sqrt{2} \sum_{n=1} \left(Y^i_n \cos n\s + \frac{i}{n}
\theta^{ij} K_{jn} \sin n\s \right) \label{str:b}
\eeq
\end{mathletters}
where $(Y^i_n, K_{in})$ are the canonical pairs. Here $\theta$ and 
$G$ are the noncommutativity parameter and the open string metric
respectively 
\begin{mathletters}
\label{theta:all}
\beq
\theta^{ij} &=& - (2\pi\a^\prime)^2
\left(\frac{1}{g+ 2\pi\a^\prime B} B\frac{1}{g- 2\pi\a^\prime B}
\right)^{ij} \label{theta:a} \\
(G^{-1})^{ij} &=& \left(\frac{1}{g+2\pi\a^\prime B}
g\frac{1}{g-2\pi\a^\prime B} \right)^{ij}. \label{theta:b}
\eeq
\end{mathletters}
We find that the open string action is written in this representation as
\beq
S_M + S_B = - \frac{1}{4\pi\a^\prime} \int_M d^2 \xi \sqrt{-h}
h^{\a\b} \left(G_{ij}  \frac{\p Y^i}{\p \xi^\a} \frac{\p Y^j}{\p \xi^\b}
+g_{ab} \frac{\p Y^a}{\p \xi^\a} \frac{\p Y^b}{\p \xi^\b} \right) 
\label{ncrep}
\eeq
where $Y^\m$ is the commutative open string variable 
\beq
Y^\m(\s) &=& x^\m + \sqrt{2} \sum_{n=1} Y^\m_n \cos n\s.
\eeq

The boundary interaction for tachyon condensation for the open string may be
written as 
\beq
S_T &=& i \int_{\s=0} d\t T(X) + i \int_{\s=\pi} d\t T(X).
\eeq
The end points of open string are given in the noncommutative theory as
\beq
X^i(0) = x^i +  \frac{\pi}{2} \theta^{ij} p_j + \sqrt{2} \sum_{n=1} Y^i_n,
\quad X^i(\pi) = x^i - \frac{\pi}{2} \theta^{ij} p_j +\sqrt{2} \sum_{n=1}
Y^i_n (-1)^n. 
\eeq
They differ from those in the commutative theory by the zero mode of the 
momentum and do not commute with each other
\beq
[X^i(0), X^j(0)]  = -i\pi \theta^{ij}, \quad
[X^i(0), X^j(\pi)]  = 0 , \quad
[X^i(\pi), X^j(\pi)] = i\pi \theta^{ij} .
\eeq

On the boundary $\p M$ if we adopt the proper-time $\hatta$ Eq.(\ref{hatta})
instead of $\t$ as the world-sheet time coordinate, the 
string variables are written on $\p M$ as 
\beq
X^i|_{\p M} = x^i - \frac{\pi}{2} \theta^{ij} p_j + Y|_{\p M}.
\eeq
Thus, we may write the boundary interaction of the tachyon 
condensation as
\beq
S_T = i \int_{\p M} d\hatta T(\zeta + Y), \quad 
[ \zeta^i, \zeta^j ] = i \pi \theta^{ij}. 
\eeq
It can be read in the closed string world-sheet coordinates as
\beq
S_T = i \int_{\p M} d\s T(\zeta + Y),
\eeq
where $Y^i$ is the usual closed string variable which can be 
expanded on $\p M$ as 
\beq
Y^i(\s) &=& y^i_0 + \sqrt{\frac{\a^\prime}{2}}
\sum_{n=1} \frac{1}{\sqrt{n}}\left(y^i_n e^{2in\s} 
+ \bary^i_n e^{-2in\s}\right). \label{y} 
\eeq 
When we transcribe the open string representation into the closed
string representation we keep noncommutative zero mode part 
unchanged, i.e., we treat $\zeta$ as noncommutative operators.
For the sake of simplicity we only consider hereafter the case where 
$Dp$-brane $\rightarrow$ $D(p-2)$-brane. Extension to the more 
general cases is straightforward. So we take 
$B_{ij} = u_{ij} = 0$ for $i, j = 0, p-2$ and $B_{ij} \not= 0$, 
$u_{ij} \not=0$ for $i,j= p-1, p$. It is convenient to introduce 
`creation' and `annihilation' operators as
\beq
b &=& \frac{1}{\sqrt{2\pi\theta}}(\zeta^{p-2} + i\zeta^{p-1}), \quad
b^\dagger = \frac{1}{\sqrt{2\pi\theta}}(\zeta^{p-2}- i\zeta^{p-1}), \quad
[b, b^\dagger] = 1. 
\eeq
We can introduce the creation and annihilation operators similarly
also for higher $Dp$-branes as we cast $(\theta)$ into the 
standard skew-diagonal form, $\theta^{2i-1,2i} = \theta^i$
\beq
(\theta) = \left( \begin{array}{ccccc}
0       & \theta_1 &        &  & \\
-\theta_1 & 0      &        &  & \\
        &          & \ddots &  & \\
        &          &        & 0 & \theta_{\frac{p}{2}} \\
        &          &        & -\theta_{\frac{p}{2}} & 0 
        \end{array} \right). \nn
\eeq

The excitations in the zero mode can be easily described as 
we introduce a complete set of the number eigenstates $\{ |n \rangle_{NC} \}$ 
\beq
\vert n \rangle_{NC} = \frac{(b^\dagger)^n}{\sqrt{n!}} \vert 0 \rangle_{NC},
\quad b \vert 0 \rangle_{NC} = 0 . \nn
\eeq
Thus, the quantum state on $\p M$, can be specified by $|n \rangle_{NC} 
\otimes | Y \rangle$, where 
\beq
\hat{Y}^i | Y \rangle = Y^i | Y \rangle. \nn
\eeq
We may expand $\hatY^i$ on $\p M$ as 
\beq
\hatY^i (\s) = \haty^i + \sqrt{\frac{\a^\prime}{2}} \sum_{n=1} 
\frac{1}{\sqrt{n}} \left\{\left(a^i_n+ \tilde{a}^{i\dagger}_n\right) 
e^{2in\s}+ \left(a^{i\dagger}_n + 
\tilde{a}^i_n \right)e^{-2in\s} \right\}, \label{hatY} 
\eeq
where $i, j = p-1, p$,
\beq
[ a^i_n, a^{j \dagger}_m ] &=& (G^{-1})^{ij} \delta_{nm}, \quad
[ \tilde{a}^i_n, \tilde{a}^{j \dagger}_m ] = (G^{-1})^{ij} 
\delta_{nm}. \label{aa}
\eeq
Since the action for
the higher modes are identical to that in the absence of the NS $B$-field
except for the space-time metric $g$ being replaced by the open string metric
$G$, we find that the boundary action in the closed string representation is 
given in the noncommutative theory as 
\beq
S_T = i\pi \zeta^i u_{ij} \zeta^j + i\pi \a^\prime \sum_{n=1}
\frac{1}{n} \bary^i_n u_{ij} y^j_n.
\eeq
From the analysis of the noncommutative open string theory, 
we may define the boundary state in the noncommutative theory as
\beq
|B_T \rangle_{NC} = \frac{T_p}{G_s} \int D[y, \bar{y}] 
\,\, \sqrt{\det(2\pi\theta)}\, {\rm tr} \,\left(e^{iS_T[y,{\bar y}]}\right) 
|y, {\bar y} \rangle
\eeq
where $G_s$ is the string coupling constant in the noncommutative 
theory. The integration measure $D[y,\bar{y}]$ does not take the zero 
modes into account and the integration over the zero modes is taken care 
of by $\sqrt{\det(2\pi\theta)}\, {\rm tr}$.
By a simple algebra we find 
\begin{mathletters}
\beq
|B_T \rangle_{NC} &=& Z \exp \left\{a^{i\dagger}_n g_{ij} 
\tilde{a}^{j\dagger}_n + a^{k\dagger}_n (G{\cal M})_{kl}
\tilde{a}^{l\dagger}_n - a^{a\dagger}_n g_{ab} 
\tilde{a}^{b\dagger}_n \right\} |0 \rangle, \\
{\cal M} &=& \left(G+ \frac{\2pap}{n}u \right)^{-1}
\left(G- \frac{\2pap}{n}u \right) \\
Z &=& \frac{T_p}{G_s} \sqrt{\det(2\pi\theta)}\,{\rm tr} 
\left( e^{- 2\pi \zeta^i u_{ij} \zeta^j} \right) 
\prod_{n=1} \det\left(G+ \frac{\2pap}{n}u \right)^{-1} 
\eeq
\end{mathletters}
where $i, j = 0, \dots, p-2$, $k, l = p-1, p$ and $a, b = p+1, 
\dots, d-1$.
Let us take the large $B$-field limit where 
\beq
\2pap B \rightarrow \infty
\eeq
with $g$ kept fixed, or equivalently the decoupling limit, where
\beq
g \sim \epsilon, \quad \a^\prime \sim \sqrt{\epsilon}, \quad
\epsilon \rightarrow 0, \label{swlim}
\eeq
while $G$, $\theta$ are kept fixed.
Note that in this limit 
\beq
\theta \rightarrow \frac{1}{B}, \quad 
G^{-1} \rightarrow - \frac{1}{(\2pap)^2} \frac{1}{B} g 
\frac{1}{B} \label{dec}
\eeq
and the effect of the tachyon condensation on the higher modes of 
$Y^i$ are suppressed. Thus, in the large $B$-field limit, we have
\beq
Z = \frac{T_p}{G_s} \sqrt{\det(2\pi\theta)}\,
{\rm tr} \left( e^{- 2\pi \zeta^i u_{ij} \zeta^j} \right) \sqrt{\det G}. 
\eeq
Now let us suppose that the system reaches the infrared fixed 
point where $u \rightarrow \infty$,
\beq
{\rm tr} \left(e^{- 2\pi \zeta^i u_{ij} \zeta^j} \right) = \sum_{n=0} 
e^{-2\pi^2 \theta (u_{p-1,p-1} + u_{pp})n} \rightarrow 1.
\eeq
It implies that the unstable D-brane behaves like a noncommutative soliton
\cite{gms} and the most symmetric one $|0 \rangle_{NC} \langle 0 |_{NC}$ 
is singled out at the infrared fixed point. Thus, at the infrared fixed point, 
we find
\beq
Z = \frac{T_p}{G_s} \sqrt{\det(2\pi\theta)} \,\sqrt{\det G}.
\eeq
We may recall the relationship between the string coupling in the 
commutative theory and that in the noncommutative one \cite{seiberg,tlee} 
\beq
\frac{G_s}{g_s} = \left(\frac{\det G}{\det g} 
\right)^{\frac{1}{4}} = \frac{\sqrt{\det G}}{\sqrt{\det(g+ \2pap B)}},
\eeq
which implies in the large $B$-field limit
\beq
\frac{G_s}{g_s} = \frac{\sqrt{\det G}}{\sqrt{\det(\2pap B)}}.
\eeq
Then it follows from Eq.(\ref{dec}) that
\beq
Z = (2\pi)^2 \a^\prime \frac{T_p}{g_s}.
\eeq
Since it can be identified with
\beq
Z = \frac{T_{p-2}}{g_s}, \nn
\eeq 
we obtain the relationship between the tension of $Dp$-brane and
that of $D(p-2)$-brane
\beq
T_{p-2} = (2\pi)^2 \a^\prime T_p.
\eeq
In the noncommutative theory the unstable D-brane is described by 
the noncommutative soliton, which corresponds to 
\beq
\phi_0(r) = 2 e^{-r^2/\sqrt{\det \theta}}.
\eeq
In the large B-field limit, the soliton becomes sharply 
localized but its contribution to the partition function is finite
in contrast to the commutative case. So the cancellation observed 
in the commutative theory is no longer needed. In fact the 
contributions from the higher modes to the partition function
are rather trivial in the noncommutative theory.
It should be noted that if we reverse the limit procedure, i.e.,
take the limit, $u \rightarrow \infty$ first then the large 
B-field limit later, we cannot get the correct result.

Now let us call our attention to the quantum state of the system
$|B_T \rangle$, which has never been discussed explicitly
in the literature. If we take the large B-field limit first, 
the boundary state reduces to
\beq
|B_T \rangle_{NC} = Z \prod_{n=1} \exp \left\{a^{i\dagger}_n g_{ij} 
\tilde{a}^{j\dagger}_n + a^{k\dagger}_n G_{kl}
\tilde{a}^{l\dagger}_n - a^{a\dagger}_n g_{ab} 
\tilde{a}^{b\dagger}_n \right\} |0 \rangle \label{bt}
\eeq
where $i, j = 0, \dots, p-2$, $k, l = p-1, p$ and $a, b = p+1, 
\dots, d-1$.
To our surprise, the resultant boundary state satisfies the 
Neumann boundary condition along the directions, $p-1, p$. It seems that
the boundary state still describes the $Dp$-brane instead of $D(p-2)$-brane.
It is certainly against our expectation. We may be tempted to take
the limit of the infrared fixed point, $u \rightarrow \infty$
prior to the large B-field limit to get the boundary state satisfying 
the Dirichlet boundary condition as in the case of the commutative theory. 
But then as we pointed out, we cannot obtain the 
correct partition function. The resolution of this problem can be 
found if we take notice that the open string metric $G$ becomes 
singular in the large B-field limit and the left movers and the 
right movers ($a$, $a^\dagger$, $\tilde{a}$, $\tilde{a}^\dagger$)
in Eq.(\ref{bt}) respect the open string metric $G$ as in 
Eq.(\ref{aa}). In order to compare the boundary state $|B_T 
\rangle_{NC}$ in the noncommutative theory with that in the commutative 
theory $|B_{B+T} \rangle$, we should rewrite the left and right 
movers in Eq.(\ref{bt}) in terms of the left and right movers in 
the commutative theory, which respect the closed string metric 
$g$. 

If we define the background metric $E$ as
\beq
E = E_S + E_A = g + \2pap B,
\eeq
equivalently,
\beq
g = \half(E + E^T), \quad B = \frac{1}{\2pap} \half (E - E^T),
\eeq
the string action with the NS B-field may be written in the closed
string world-sheet coordinates as
\beq
S_M + S_B = -\frac{1}{4\pi\a^\prime} \int_M d^2 \xi \sqrt{-h} h^{\a\b}
E_{ij} \frac{\p X^i}{\p \xi^\a} \frac{\p X^j}{\p \xi^\b}. \label{actcm}
\eeq
Here we are concerned with the action only for the string 
variables in the directions of $p-1, p$. If we apply the
open-closed string duality to Eq.(\ref{ncrep}), we obtain the 
string action in the noncommutative theory as 
\beq
S_M + S_B = -\frac{1}{4\pi\a^\prime} \int_M d^2 \xi \sqrt{-h} h^{\a\b} G_{ij}
\frac{\p Y^i}{\p \xi^\a} \frac{\p Y^j}{\p \xi^\b}. \label{actnc}
\eeq
Comparing the string action in the commutative theory Eq.(\ref{actcm}) 
with that in the noncommutative theory Eq.(\ref{actnc}), we find 
that two actions are related by the well-known T-dual 
transformation \cite{giveon}
\beq
E^\prime = (aE + b) (cE + d)^{-1} 
\eeq
where $a$, $b$, $c$ and $d$ satisfy the following $O(2,2,R)$ 
condition
\beq
\left(\begin{array}{cc} 
  a & b \\
  c & d  
\end{array}\right)^T 
\left(\begin{array}{cc}
  0 & I \\
  I & 0 
\end{array}\right) 
\left(\begin{array}{cc}
  a & b \\
  c & d  
\end{array}\right) &=& 
\left(\begin{array}{cc}
  0 & I \\
  I & 0  
\end{array}\right).
\eeq
Under the T-dual transformation the left and right movers 
transform as
\begin{mathletters}
\beq
a_n(E) & \rightarrow & (d- cE^T)^{-1} a_n(E^\prime),\quad
a^\dagger_n(E) \rightarrow a^\dagger_n (E^\prime)(d^T -Ec^T)^{-1}, \\
\tilde{a}_n(E) & \rightarrow & (d+ cE)^{-1}\tilde{a}_n(E^\prime),\quad
\tilde{a}^\dagger_n(E) \rightarrow \tilde{a}^\dagger_n (E^\prime)
(d^T +E^Tc^T)^{-1} 
\eeq
\end{mathletters}
Choosing $E^\prime = G$, we find the T-dual transformation \cite{tlee2}, 
which connects the left movers and the right movers in the commutative 
theory Eq.(\ref{actcm}) and those in the noncommutative theory 
Eq.(\ref{actnc}) 
\beq
T = \left(\begin{array}{cc}
  a & b \\
  c & d 
\end{array}\right)
=  \left(\begin{array}{cc}
  I & 0 \\
 - (\2pap)^{-1} \theta & I
\end{array}\right).
\eeq
If we denote the left and right movers in the commutative theory, 
Eq.(\ref{actcm}) by $a$, $a^\dagger$, $\tilde{a}$, 
$\tilde{a}^\dagger$, and those in the noncommutative theory, Eq.(\ref{actnc})
by $a^\prime$, $a^{\dagger\prime}$, $\tilde{a}^\prime$, 
$\tilde{a}^{\dagger\prime}$, the relationship between them are 
given by
\begin{mathletters}
\beq
a_n^\prime &=& \left(I+\frac{\theta E^T}{\2pap} \right) a_n, \quad
a^{\dagger\prime}_n = a^{\dagger}_n \left(I-\frac{E \theta}{\2pap}\right), \\
\tilde{a}^\prime_n &=& \left(I-\frac{\theta E}{\2pap} \right) 
\tilde{a}_n, \quad
\tilde{a}^{\dagger\prime}_n = \tilde{a}^{\dagger}_n \left(I+ \frac{
E^T \theta}{\2pap}\right).
\eeq
\end{mathletters}
Note that
\beq
I - \frac{E \theta }{\2pap} = EG^{-1}, \quad
I + \frac{E^T \theta}{\2pap} = E^T G^{-1}. \label{int1}
\eeq
Using Eq.(\ref{int1}) and Eq.(\ref{dec}) in the large B-field 
limit, we obtain
\beq
\exp \left(a^{\dagger\prime}_n
G \tilde{a}^{\dagger\prime}_n \right)|0 \rangle
= \exp\left(a^{\dagger}_n E G^{-1} E \tilde{a}^{\dagger}_n \right)
|0 \rangle = \exp \left(-a^{\dagger}_n
g \tilde{a}^{\dagger}_n \right)|0 \rangle.
\eeq
Therefore, the boundary state turns out to satisfy the the correct Dirichlet 
boundary condition along the directions of $p-1, p$
\beq
|B_T \rangle_{NC} = Z \prod_{n=1} \exp \left\{a^{i\dagger}_n g_{ij} 
\tilde{a}^{j\dagger}_n - a^{a\dagger}_n g_{ab} 
\tilde{a}^{b\dagger}_n \right\} |0 \rangle \label{btt}
\eeq
where $i, j = 0, \dots, p-2$ and $a, b = p-1, \dots, d-1$.
In the noncommutative theory 
the boundary state also describes the $D(p-2)$-brane as desired.
One can reach the same conclusion also by taking the Seiberg-Witten limit 
Eq.(\ref{swlim}).
One may attempt to get the same result in the commutative theory, 
by taking the large B-field limit first. As one may expect,
\beq
|B_{F+T} \rangle \rightarrow Z_{Disk} \prod_{n=1} \exp \left\{a^{i\dagger}_n 
g_{ij} \tilde{a}^{j\dagger}_n - a^{a\dagger}_n g_{ab} 
\tilde{a}^{b\dagger}_n \right\} |0 \rangle
\eeq
where $i, j = 0, \dots, p-2$ and $a, b = p-1, \dots, d-1$.
However, in this limit $Z_{Disk}$ does not yield the correct 
tension of the D-brane. Hence, this procedure only works for the 
noncommutative theory.
In order to get a consistent description of 
tachyon condensation in the noncommutative theory, we should take the 
large B-field limit prior to the infrared fixed point limit. 
It is also consistent with the work of Gopakumar, Minwalla and Strominger
\cite{gms}.

\section{Discussions and Conclusions}

In this paper we discussed the tachyon condensation, which is one of the noble 
phenomena in string theory, in the simplest setting, i.e., in a single D-brane 
system in the bosonic string theory, using the boundary state formulation.
We do not need to deal with the non-Abelian supersymmetric 
formulation to understand the tachyon condensation in this 
simplest system unlike in the $D$-${\bar D}$-brane system. 
For the purpose of applying the boundary state formulation to the system with
boundary actions of general form, we improved the boundary state 
formulation of ref.\cite{callan}. We show that in general the 
normalization factor of the boundary state corresponds to the disk 
partition function for the given boundary interaction and obtain 
the general form of the boundary state. As the tachyon 
condensation develops the boundary state wavefunction becomes 
sharply localized in the directions where the tachyon condensation 
occurs and eventually reduces to that of a lower dimensional D-brane. 
Both tension and boundary state wavefunction of the lower dimensional 
D-brane are correctly derived from the boundary state formulation 
of the unstable D-brane at the infrared fixed point. Since one may 
obtain the non-BPS $D2p$-brane ($D(2p+1)$-brane) of type IIB (IIA) string 
theory, starting from a $D2p$-${\bar D}2p$-brane 
($D(2p+1)$-${\bar D}(2p+1)$-brane) pair in type IIA (IIB) string 
theory \cite{sen1}, it is interesting to discuss these descent 
relations among BPS and the non-BPS $D$-brane in the boundary state formulation, 
extending the present work. The relationship between the present work and those 
on the $D$-${\bar D}$ system \cite{kuta00,kraus00,alwis} may be clarified in 
this context. 

As we point out that the boundary state formulation, discussed in 
this paper, has some advantages over other approaches, in that the
it provides not only the disk partition function but also the 
quantum state of the system. Since the boundary state depicts the system in
terms of the quantum state of the closed string and the disk partition 
function corresponds to the normalization of the closed string 
wavefunction, it may be possible to embed it in a large framework, 
the closed string field theory. The open-closed string duality then may 
lead us to the open string field theory of the tachyon condensation. 
It may improve our understanding of the open string field theory of 
tachyon condensation, which has been discussed only within the limits 
of the level truncation \cite{senzw}. 

One of the main results of the present work is that the noncommutative 
theory of the tachyon condensation is derived from the 
boundary state formulation and its relationship to the commutative 
theory is established along the line of the equivalence between 
the noncommutative open string theory and the noncommutative one 
in canonical quantization. We show that in the noncommutative theory 
the unstable D-brane precisely reduces to the lower dimensional D-brane 
in the large B-field. The boundary state formulation explains why some 
of the results obtained in the large $B$-field limit in the noncommutative 
theory are exact. The zero mode contribution is the most important 
and the contributions of other higher modes are suppressed.
However, there is a subtle point, yet important that at its appearance 
the resultant boundary state of the unstable D-brane satisfies the Neumann 
boundary condition in stead of the Dirichlet boundary condition 
along the directions where the tachyon condensation develops.
But taking notice that the left and the right movers along the 
directions of the tachyon condensation respect the open string 
metric $G$, which is singular in the large B-field limit, we find 
a resolution in the framework of the boundary state formulation.
Using the T-dual relationship between the left and right movers in 
the commutative theory and those in the noncommutative theory,
we find that the boundary state correctly reduces to that of the lower 
dimensional D-brane. 

We may take a step forward in the noncommutative theory
by introducing the $U(1)$ background in addition to the 
NS B-field background. The extension along this direction is 
important in connection with the confinement in $D$-${\bar D}$ 
system \cite{berg} and the matrix model description of the tachyon
condensation \cite{matrix}. It may be also fruitful to extend the present 
work along other directions, such as the quantum corrections at the one 
loop level to the tachyon condensation \cite{oneloop,itoyama} and the tachyon 
condensation on noncommutative tori \cite{torus}, which have been discussed 
recently in the literature.

\section*{Acknowledgement}
This work was supported by grant No. 2000-2-11100-002-5 from the Basic Research 
Program of the Korea Science \& Engineering Foundation. Part of this work was 
done during the author's visit to APCTP (Korea), KIAS (Korea) and PIMS (Canada). 
The author thanks G. Semenoff for his contribution at the 
early stage of this work and for hospitality during the author's
visit to PIMS. He also thanks Piljin Yi, S. J. Rey and S. Hyun
for useful discussions.


\begin{thebibliography}{99}

\bibitem{sen1} A. Sen, JHEP 9808 (1998) 012, [hep-th/9805170], Tachyon condensation 
on the brane antibrane system; Int. Jour. Mod. Phys. A14 (1999) 4061, 
[hep-th/9902105], Decent relations among bosonic D-branes; APCTP Winter school 
lecture, [hep-th/9904207], Non-BPS states and branes in string theory;
JHEP 9912 (1999) 027, [hep-th/9911116], Universality of the tachyon potential.

\bibitem{tach} For early works on this subject see, K. Bardakci, Nucl. Phys.
B68 (1974) 331, Dual Models and Spontaneous Symmetry Breaking;
K. Bardakci and M. B. Halpern, Phys. Rev. D10 (1974) 4230,
Explicit Spontaneous Breakdown in a Dual Model; Nucl. Phys. 
B96 (1975) 285, Explicit Spontaneous Breakdown in a Dual Model II: N Point Functions; 
K. Bardakci, Nucl. Phys. B133 (1978) 297, 
Spontaneous Symmetry Breakdown in the Standard Dual String Model. 

\bibitem{witt} E. Witten, Nucl. Phys. B268 (1986) 253, 
Noncommutative geometry and string field theory.

\bibitem{kost} V. A. Kostelecky and S. Samuel, Phys. Lett. B207 
(1988) 169, The static tachyon potential in the open bosonic 
string theory; Nucl. Phys. B336 (1990) 263,
On a nonperturbative vacuum for the open string.

\bibitem{senzw}
A. Sen and B. Zwiebach, JHEP 0003 (2000) 002, [hep-th/9912249], 
Tachyon condensation in string field theory;
N. Moeller and W. Taylor, Nucl. Phys. B583 (2000) 105, 
[hep-th/0002237], Level truncation and the tachyon in open bosonic 
string field theory; 
A. Kostelecky and R. Potting, [hep-th/0008252], Analytical 
construction of a nonperturbative vacuum for the open bosonic string;
B. Zwiebach, [hep-th/0010190], Trimming the tachyon string field with 
$SU(1,1)$.

\bibitem{witt92}
E. Witten, Phys. Rev. D46 (1992) 5467, [hep-th/9208027], On 
background independent open string field theory; Phys. Rev. D47 (1993) 3405, 
[hep-th/9210065], Some computations in background of independent off-shell 
string theory.

\bibitem{shata} S. L. Shatashvili, Phys. Lett. B311 (1993) 83, 
[hep-th/9303143], Comment on the background independent open 
string theory; [hep-th/9311177], On the problems with background 
independence in string theory; A. A. Gerasimov and S. L. 
Shatashvili, [hep-th/0009103], On exact tachyon potential in open 
string field theory.

\bibitem{sfttr} A. Sen and B. Zwiebach, JHEP 0010 (2000) 009
[hep-th/0007153] Large marginal deformations in string field 
theory; J. A. Harvey and P. Kraus, JHEP 0004 (2000) 012, [hep-th/0002117],
D-branes as unstable lumps in bosonic open string field theory;
R. de Mello Koch, A. Jevicki, M. Mihailescu and R. Tatar, Phys. 
Lett. B482 (2000) 249 [hep-th/0003031], Lumps and p-branes in open 
string field theory;
N. Moeller, A. Sen and B. Zwiebach, [hep-th/0005036], D-branes as 
tachyon lumps in string field theory;
R. de Moeller Koch and J. P. Rodrigues, [hep-th/0008053], Lumps in 
level truncated open string field theory; 
N. Moeller, [hep-th/0008101], Codimension two lump solutions in 
string field theory and tachyonic theories;
H. Hata and S. Shinohara, JHEP 0009 (2000) 035 [hep-th/0009105],
BRST invariance of the non-perturbative vacuum in bosonic open string field
theory; M. Schnabl, [hep-th/0011238], Constraints on the tachyon condensate from 
anomalous symmetries. 

\bibitem{eff} A. A. Gerasimov and S. L. Shatashvili, 
[hep-th/0009103], On exact tachyon potential in open string field 
theory; D. Kutasov, M. Mari\~{n}o and G. Moore, 
[hep-th/0009148], Some exact results on tachyon condensation in 
string field theory.

\bibitem{witten00} E. Witten, [hep-th/0006071], Noncommutative 
tachyons and string field theory.

\bibitem{dasgupta} K. Dasgupta, S. Mukhi and G. Rajeshi, JHEP 0006 
(2000) 022, [hep-th/0005006], Noncommutative tachyons.

\bibitem{corbalba} L. Cornalba, [hep-th/0010021], Tachyon 
condensation in large magnetic fields with background independent 
string field theory.

\bibitem{oku} K. Okuyama, [hep-th/0010028], Noncommutative tachyon 
from background independent open string field theory.

\bibitem{schnabl}, M. Schnabl, [hep-th/0020034], String field theory
at large B-field and noncommutative geometry.

\bibitem{harvey00} J. A. Harvey, P. Kraus, F. Larsen and E. 
Martinec, [hep-th/0005031], D-branes and strings as noncommutative 
solitons; J. A. Harvey, P. Kraus and F. Larsen, [hep-th/0010060],
Exact noncommutative solitons. 

\bibitem{kuta00} D. Kutasov, M. Mari\~{n}o and G. Moore, 
[hep-th/0010108], Remarks on tachyon condensation in superstring 
field theory.

\bibitem{marino} M. Marino, [hep-th/0103089], On the BV 
formulation of boundary superstring field theory.

\bibitem{niarchos} V. Niarchos and N. Prezas, [hep-th/0103102],
Boundary superstring field theory.

\bibitem{kraus00} P. Kraus and F. Larsen, [hep-th/0012198], Boundary 
string field theory of the $D{\bar D}$ system;
Tadashi Takayanagi, Seiji Terashima, Tadaoki Uesugi, JHEP 0103 (2001) 
019, [hep-th/0012210], Brane-Antibrane Action from Boundary String Field 
Theory.

\bibitem{callan} C. G. Callan, C. Lovelace, C. R. Nappi and S. A. 
Yost, Nucl. Phys. B308 (1988) 221, Loop corrections to superstring 
equations of motion.

\bibitem{seiberg} N. Seiberg and E. Witten, JHEP 9909 (1999) 032, 
[hep-th/9908142], String theory and noncommutative geometry.

\bibitem{tlee} T. Lee, Phys. Rev. D62 (2000) 024022, 
[hep-th/9911140], Canonical quantization of open string and 
noncommutative geometry;
Phys. Lett. B478 (2000) 313, [hep-th/9912038], Noncommutative 
Dirac-Born-Infeld action for D-brane.

\bibitem{gsw} In the present paper, we follow the notation of 
``Superstring theory I and II" by M. B. Green, J. H. Schwarz and E. Witten,
Cambridge University Press (1987).

\bibitem{frad} E. S. Fradkin and A. A. Tseytlin, Phys. Lett. B163 
(1985) 123, Nonlinear electrodynamics from quantized strings;
J. Ambjorn, Y. M. Makeenko, G. W. Semenoff and R. Szabo, 
[hep-th/0012092], String theory in electromagnetic fields.

\bibitem{vecch} P. Di Vecchia, M. Frau, A. Lerda and A. Liccardo,
[hep-th/9906214], $(F, Dp)$ bound states from the boundary state;
P. Di Vecchia and A. Liccardo, hep-th/9912275, D branes in string 
theories II.

\bibitem{andreev} O. Andreev, [hep-th/0010218], Some computations 
of partition functions and tachyon potentials in background 
independent off-shell string theory.

\bibitem{alwis}
S. P. de Alwis, [hep-th/0101200], Boundary string 
field theory, the boundary state formalism and D-brane tension.

\bibitem{semenoff} 
E. T. Akhemedov, M. Laidlaw and G. W. Semenoff,
[hep-th/0106033], On a modinifaction of the boundary state 
formalism in off-shell string theory.

\bibitem{gms} R. Gopakumar, S. Minwalla and A. Strominger, JHEP 05 
(2000) 020, [hep-th/0003160], Noncommutative solitons.

\bibitem{giveon} A. Giveon, M. Porrati and E. Rabinovici, Phys. 
Rep. 244 (1994) 77, [hep-th/9401139], 
Target space duality in string theory.

\bibitem{tlee2} T. Lee, Phys. Lett. B483 (2000) 277, 
[hep-th/0004159], Open superstring and noncommutative geometry.
 
\bibitem{gms2} R. Gopakumar, S. Minwalla and A. Strominger,
[hep-th/0007226], Symmetry restoration and tachyon condensation in 
open string theory.

\bibitem{berg} O. Bergman, K. Hori and Piljin Yi, [hep-th/0002223], 
Confinement on the Brane.

\bibitem{matrix} N. Seiberg, [hep-th/0008013], A note on 
background independence in noncommutative gauge theories, matrix 
model and tachyon condensation; M. Li, [hep-th/0010058], Note on noncommutative tachyon in
matrix models; G. Mandal and S. R. Wadia, [hep-th/0011094], Matrix model, 
noncommutative gauge theory and the tachyon potential.

\bibitem{oneloop} T. Suyama, [hep-th/0102192], Tachyon 
condensation and spectrum of strings on D-branes;
K. S. Viswanathan and Y. Yang, [hep-th/0104099], Tachyon 
condensation and background independent superstring field theory;
M. Alishahiha, [hep-th/0104164], One-loop correction of the
tachyon action in boundary superstring field theory; 
K. Bardakci and A. Konechny, [hep-th/0105098],
Tachyon condensation in boundary string field theory at one loop;
B. Craps, P. Kraus, F. Larsen, JHEP 0106 (2001) 062, [hep-th/0105227], 
Loop Corrected Tachyon Condensation.

\bibitem{itoyama} A. Fujii and H. Itoyama, [hep-th/0105247], Some 
computation on $g$-function and disc parition function and 
boundary string field theory. 

\bibitem{torus} I. Bars, H. Kajiura, Y. Matsuo, T. Takayanagi,
Phys.Rev. D63 (2001) 086001 [hep-th/0010101], Tachyon Condensation on 
Noncommutative Torus; H. Kajiura, Y. Matsuo and T. Takayanagi,
[hep-th/0104143], Exact tachyon condensation on noncommutative 
torus.











\end{thebibliography}
\end{document}